\documentclass[12pt,a4paper]{article}
\pdfoutput=1
\usepackage{jheppub}
\usepackage{amsfonts}
\usepackage{bbm}
\usepackage{graphicx}
\usepackage{caption}
\usepackage{subcaption}
\usepackage[utf8]{inputenc}
\usepackage[T1]{fontenc}
\usepackage{amssymb,amsfonts,amsmath}
\usepackage{epsfig}

\font\cmss=cmss10
\font\cmsss=cmss10 at 7pt
\font\manual=manfnt

\newcommand{\bi}{\begin{itemize}}
\newcommand{\ei}{\end{itemize}}

\newcommand{\bea}{\begin{eqnarray}}
\newcommand{\eea}{\end{eqnarray}}
\newcommand{\be}{\begin{equation}}
\newcommand{\ee}{\end{equation}}
\newcommand{\ben}{\begin{eqnarray*}}
\newcommand{\een}{\end{eqnarray*}}
\newcommand{\bem}{\begin{pmatrix}}
\newcommand{\eem}{\end{pmatrix}}
\newcommand{\bl}{\begin{align}}
\newcommand{\el}{\end{align}}
\newcommand{\beg}{\begin{gather}}
\newcommand{\eeg}{\end{gather}}

\newenvironment{myitemize}{
\begin{itemize}
   \setlength{\itemsep}{1pt}
   \setlength{\parskip}{0pt}
   \setlength{\parsep}{0pt}}{\end{itemize}}
   
\newcommand{\cA}{\mathcal{A}}

\newcommand{\cO}{\mathcal{O}}

\newcommand{\IH}{\mathbb{H}}

\renewcommand{\a}{\alpha}
\renewcommand{\b}{\beta}
\renewcommand{\d}{\delta}
\newcommand{\e}{\epsilon}
               
\newcommand{\g}{\gamma}
\newcommand{\h}{\eta}

\renewcommand{\k}{\kappa}             
\renewcommand{\l}{\lambda}
\newcommand{\m}{\mu}
\newcommand{\n}{\nu}

\renewcommand{\r}{\rho}                                     
\newcommand{\s}{\sigma}                                   
\renewcommand{\t}{\tau}

\newcommand{\D}{\Delta}

\newcommand{\G}{\Gamma}

\renewcommand{\L}{\Lambda}
\renewcommand{\O}{\Omega}

\newcommand{\vf}{\varphi}
\newcommand{\ve}{\varepsilon}

\newcommand{\half}{\frac{1}{2}}

\newcommand{\nn}{\nonumber}

\def\dbend{\lower3.5pt\hbox{\manual\char127}}

\def\IL{\relax{\rm I\kern-.18em L}}
\def\IH{\relax{\rm I\kern-.18em H}}
\def\rlx{\relax\leavevmode}

\def\ZZ{\rlx\leavevmode\ifmmode\mathchoice{\hbox{\cmss Z\kern-.4em Z}}
 {\hbox{\cmss Z\kern-.4em Z}}{\lower.9pt\hbox{\cmsss Z\kern-.36em Z}}
 {\lower1.2pt\hbox{\cmsss Z\kern-.36em Z}}\else{\cmss Z\kern-.4em
 Z}\fi}

\title{Quantum Cosmology Near Two Dimensions}

\author[1, 2]{Teresa Bautista}
\author[1, 2, 3]{and Atish Dabholkar}

\affiliation[1]{Sorbonne Universit\'es, UPMC Univ Paris 06\\
  UMR 7589, LPTHE, F-75005, Paris, France}

\affiliation[2]{CNRS, UMR 7589, LPTHE, F-75005, Paris, France}

\affiliation[3]{International Centre for Theoretical Physics\\
ICTP-UNESCO, Strada Costiera 11, Trieste 34151 Italy}

\abstract{We consider a Weyl-invariant formulation of gravity with a cosmological constant in  $d$-dimensional spacetime and show that near two dimensions the classical action reduces to the timelike Liouville action.  We show  that the  renormalized  cosmological term leads to a nonlocal quantum momentum tensor  which satisfies the Ward identities in a nontrivial way.  The resulting   evolution equations for an isotropic, homogeneous universe lead  to  a slowly decaying vacuum energy and a power-law expansion. We outline the implications for   the cosmological constant problem,  inflation, and dark energy.}

\keywords{Liouville Theory, Cosmological Constant, Inflation, Dark Energy}

\begin{document}
\maketitle

\section{Introduction}

It was pointed out recently \cite{Dabholkar:2015qhk, Dabholkar:2015b} that the inclusion of Weyl anomalies can have striking consequences for the gravitational dynamics on cosmological scales. In particular, it offers a new perspective on inflation, dark energy, and the cosmological constant problem. To  deal with these anomalies it is convenient to use a Weyl-invariant formulation of gravity \cite{Dabholkar:2015qhk, Dabholkar:2015b}.  
In this paper we analyze  this formalism and the consequences of Weyl anomalies near two dimensions.

The  Weyl invariant formulation is obtained by introducing  a Weyl compensator field and a fiducial metric  which scale appropriately keeping the physical metric Weyl invariant. The resulting theory has an extra scalar degree of freedom but also an enlarged   gauge symmetry which  includes Weyl symmetry in addition to diffeomorphisms.  The number of degrees of freedom remains the same  upon imposing Weyl invariance. See \cite{Zumino:1970tu,Deser:1970hs,Kaku:1977pa,Das:1978nr,Gover:2008sw,Duff:1993wm,Iorio:1996ad,Jackiw:2005su,tHooft:2011aa,Codello:2012sn} for related earlier work.  

This formalism  is convenient for studying  the renormalization of the quantum gravity path integral because it  separates general coordinate invariance from Weyl invariance.  On general grounds, Weyl invariance can have anomalies in the quantum theory  because renormalization  introduces a scale.  An essential requirement of the enlarged gauge principle is that  all such potential anomalies cancel because Weyl symmetry is a gauge symmetry.  General coordinate invariance of the original theory then becomes  equivalent to general coordinate invariance plus  \textit{quantum} Weyl invariance of the modified theory. This provides a useful guiding principle.

In two spacetime dimensions, the Weyl-invariant formulation is particularly advantageous because the  metric tensor has only three components. With the enlarged gauge symmetry, one can choose a gauge in which the fiducial metric is completely fixed, with no dynamics.  The entire quantum dynamics then resides in the dynamics of the  scalar Weyl compensator in the  background of the fixed fiducial metric. 
Moreover, the classical action for the Weyl compensator  reduces to an analytic continuation of the well-studied Liouville action. 

Even with  this simplification, the quantization of two-dimensional gravity presents many difficulties because the path integral measure is not shift-invariant and the cosmological constant term  is a  nontrivial  exponential interaction.  Moreover, the kinetic term for the Weyl compensator  is negative or `timelike', so  the Weyl compensator  is related to the  Liouville field by a subtle analytic continuation which is not yet fully understood in the quantum theory \cite{Strominger:2003fn, Schomerus:2003vv, Zamolodchikov:2005fy,Kostov:2005kk,McElgin:2007ak,Harlow:2011ny,Giribet:2011zx,Ribault:2015sxa}. As a result, the quantum  gravitational dynamics  is  highly nontrivial even in two dimensions\footnote{There  is extensive literature on both spacelike and timelike Liouville theory. See \cite{Ribault:2014hia, Nakayama:2004vk, Teschner:2001rv, Ginsparg:1993is, AlvarezGaume:1991kd,Harlow:2011ny} for reviews that emphasize  different aspects of quantum Liouville theory.}. Some of the subtleties of the nonperturbative quantization and the question of the existence  of the quantum theory are discussed in \S\ref{Existence}.

Fortunately, the full machinery of timelike Liouville theory is not necessary to address the  physical questions that we are interested in. Our main aim is to study the cosmological consequences of the  anomalous gravitational dressing of quantum operators,  especially the cosmological constant operator. We would like to use the two-dimensional model to draw some general lessons that may be applicable in four dimensions. We expect that the semiclassical approximation should be reliable on cosmological scales in four dimensions. Any effect in the two-dimensional model that could be relevant for four-dimensional physics must  manifest itself in the semiclassical limit and should not depend on special properties of two dimensions. For this reason, we  confine ourselves to the semiclassical limit of timelike Liouville theory. 

One of the surprising results we find is that  the quantum corrections lead to a slow decay of vacuum energy in an isotropic and homogeneous universe and a slowing down of the exponential de Sitter expansion.  Since the essential mechanism relies purely on the infrared physics, we expect that this two-dimensional example can  serve as an interesting model for possible generalizations to higher dimensions.   

The idea of  vacuum energy decay caused by infrared quantum effects has been explored earlier in four-dimensional gravity by several physicists\footnote{An interesting related idea explored in the literature concerns  possible nontrivial  fixed points of gravity in the UV \cite{Weinberg:1980gg,Kawai:1989yh, Kawai:1992np, Kawai:1993mb, Aida:1994zc, Bonanno:2001xi,Reuter:2005kb,Weinberg:2009wa,Bonanno:2010bt,Reuter:2012xf} and in the IR \cite{Antoniadis:1991fa,Antoniadis:1998fi}.  This is a different regime than what we consider. Our interest is in the long distance physics on cosmological scales in weakly coupled gravity near the \textit{trivial} Gaussian fixed point. Some of the  methods developed in these investigations could nevertheless be useful  for the computation of Weyl anomalies especially in very early universe.}. There is considerable divergence in the literature about the final result \cite{Polyakov:1982ug,Mottola:1984ar,Antoniadis:1985pj,Tsamis:1996qq,Polyakov:2007mm,Polyakov:2012uc,Romania:2012av} and more generally about infrared effects in nearly de Sitter spacetime \cite{Starobinsky:1994bd,Weinberg:2005vy,Weinberg:2006ac,Senatore:2009cf,Kitamoto:2010si,
Kahya:2010xh,Marolf:2010zp,Giddings:2010nc,Higuchi:2011vw,Marolf:2011sh,Pimentel:2012tw}. 
The  advantage of our two-dimensional model  is that the important quantum effects can be computed explicitly with relative ease and without ambiguities to all orders in perturbation theory. 
The main lesson  is that  the physical coupling constants are the couplings of the \textit{gravitationally dressed} operators. The anomalous dimensions of the   dressed operators are  in principle different from the anomalous dimensions of the undressed operators.  For example,  the cosmological constant, which is usually regarded as the coupling constant of the identity operator, is really the coupling constant of   the square-root of the determinant of the metric with a nontrivial anomalous gravitational dressing. This anomalous dressing introduces additional dependence on the metric and affects the gravitational dynamics.This is the essential idea that we wish to generalize to four dimensions. 

Another novelty of our approach  is that  we summarize the quantum effects in terms  of  a \textit{nonlocal} effective action for a general background metric  correctly incorporating the Weyl anomalies. The advantage of this approach is that it separates the computation of quantum effects from the analysis of the  effective dynamics that follows from the nonlocal action.  In four dimensions the quantum action can be deduced\footnote{Unlike in two dimensions, the anomalous dressing of the cosmological term cannot be computed exactly in four dimensions, but  can be computed perturbatively in a given microscopic theory \cite{Dabholkar:2015}. } as a solution to  the local renormalization group equation\cite{Dabholkar:2015b}.  The  cosmological solutions of the resulting equations of motion in four dimensions will be presented in \cite{Bautista:2015b}.  However, our  two-dimensional quantum model already captures many of the essential features in a much simpler context.   We hope that this way of organizing the calculations will prove  useful for future explorations.

The paper is organized as follows. In $\S{\ref{ClassicalGravity}}$ we summarize  relevant aspects of  the Weyl invariant formulation of classical gravity and cosmology in $d$ spacetime dimensions and show that near two dimensions the action reduces to the timelike Liouville action. 
In $\S{\ref{QuantumGravity}}$ we discuss  the renormalization of the cosmological constant operator  and compute the non-local momentum tensor which satisfies the Ward identities. 
In $\S{\ref{QuantumCosmo}}$ we study the evolution of an isotropic and homogeneous universe without matter in the presence of this nonlocal momentum tensor. We then discuss the consequences of the quantum decay of vacuum energy for the cosmological constant problem, inflation, and dark energy. 

 \section{Classical Gravity  Near Two Dimensions \label{ClassicalGravity}}
 
To obtain a  Weyl-invariant formulation of classical gravity in $d$ dimensions one  introduces a fiducial metric 
$h_{\m\n}$ and a Weyl compensator field $\O$ 
to write the physical metric as
 \begin{equation}\label{metric-relation}
g_{\m\n} = e^{2\O} h_{\m\n} \, .
\end{equation}
Given any action which is a functional of the  physical metric, one can substitute (\ref{metric-relation}) to obtain  an action as a functional of  the fiducial metric and the Weyl compensator. 
The  rewriting (\ref{metric-relation}) of the metric is invariant under a local Weyl transformation:
\be
h_{\m\n}\rightarrow e^{2\xi(x)}h_{\m\n},\qquad \O(x)\rightarrow \O(x)-\xi(x) \, .\nonumber
\ee
We call this symmetry `fiducial Weyl symmetry' to underscore the fact that even though the fiducial metric transforms under it, the physical metric is invariant.
 This somewhat trivial rewriting will help make contact with Liouville theory in two dimensions and  prepare the ground for the quantum theory.
 
 \subsection{Classical Gravity in the Weyl-invariant Formulation}

$\bullet$ \textit{Action}
\vskip2mm
\noindent The  Einstein-Hilbert gravitational action in $d$ spacetime dimensions  is given by
\begin{equation}\label{action1}
I_{K}[g] =  \frac{M_{p}^{d-2}}{2} \int d^{d}x  \, \sqrt{-g} \,  R[g]  \,  
\end{equation}
where $M_{p}$ is the reduced Planck mass and $R[g]$ is the Ricci scalar for the physical metric $g_{\m\n}$.
The cosmological constant term is given by the action
\be\label{action2}
I_{\L}[g] =  -  M_{p}^{d-2}\L \int d^{d}x \, \sqrt{-g} \, 
\ee
where  $\L$ is the cosmological constant. 
Given a  UV cutoff $M_{0}$, the Planck scale $M_{p}$ and the cosmological constant $\L$ 
correspond to  dimensionless `coupling constants' $\kappa^{2}$ and $\lambda$ defined by:
 \begin{equation}\label{dimless}
M_{p}^{d-2} := \frac{M_{0}^{d-2}}{\kappa^{2}}  \, , \qquad \Lambda :=   \lambda \kappa^{2} M_{0}^{2} \, .
\end{equation} 
Substituting (\ref{metric-relation})  in (\ref{action1}) and (\ref{action2}) we obtain  
\be\label{Om-action1}
I_{K}[\O, h]= \frac{M_{p}^{d-2}}{2}\int d^{d}x \sqrt{-h} \ e^{(d-2)\Omega} \left(  R[h]  + (d-2)(d-1) h^{\m\n}\,\nabla_{\m} \Omega \nabla_{\n} \O \right) \, 
\ee
for the gravitational action and
\be\label{Om-action2}
I_{\L}[\O, h] =  -  M_{p}^{d-2}\L \int d^{d}x  \sqrt{-h} \,   e^{d\O}
\ee
for the cosmological term. Henceforth, unless stated otherwise, all contractions and covariant derivatives are using the fiducial metric.
The actions  \eqref{Om-action1} and \eqref{Om-action2} are  invariant under the fiducial Weyl transformation by construction.
\vskip2mm
$\bullet$ \textit{Ward Identities}
\vskip1mm
\noindent The new actions (\ref{Om-action1}) and (\ref{Om-action2})  are each independently invariant under Weyl transformations in addition to coordinate transformations. As a result, they each satisfy  two Ward identities. 

Under a general coordinate transformation $x^{\mu} \rightarrow x^{\prime \m} = 
 x^{\m}  + \xi^{\m}(x) $,  a scalar  field transforms as $\O'(x') = \O(x)$ with infinitesimal diffeomorphism variation given by
 \be
   \d \O  := \O' (x) -\O(x)    = - \xi^{\m}(x) \, \nabla_{\mu} \, \O \, .\\
 \ee
 Similarly,  the variation of the metric is given by
 \bea\label{fieldcoord}
  \d h_{\m\nu} =  - \left( \nabla_{\m}\, \xi_{\n} + \nabla_{\n}\,\xi_{\m} \right)
\, , \quad   \d h^{\m\nu} =  \nabla^{\m}\, \xi^{\n} + \nabla^{\n}\,\xi^{\m} \,  . 
\eea
Invariance of the action functional implies
\be\label{diff-identity}
\d I = \int d^{d}x  \, \left[ 2 \, ( \nabla^{\n} \, \xi^{\m}(x) )\, \left( \frac{\d I}{\d h^{\m\n}} \right) - \,   \,   \xi^{\m}(x)\, \frac{\d I}{\d \O} \nabla_{\m} \O\right] \equiv 0
\ee
Consequently both $I_{K}$ and $I_{\L}$ satisfy the  Ward identities for coordinate invariance:
\be\label{Diff-WT}
\nabla^{\n} ( \,\frac{-2 \, \d I_{a}}{\sqrt{-h} \, \d h^{\m\n}})  -    \frac{1}{\sqrt{-h}} \,\frac{\d I_{a}}{\d \O}\, \nabla_{\m}\O   \,  \equiv 0 \quad (a= K, \L)\,  .
\ee
Similarly, the infinitesimal Weyl variation is given by
\be
\d h_{\m\nu} = 2\xi(x)\, h_{\m\nu} \, , \quad \d h^{\m\nu} = -2\xi(x)\,   h^{\m\nu} \, , \quad  \d\O = -\xi(x)
\ee
and the corresponding Ward identity is
\be\label{Weyl-identity}
 h ^{\m\n}  ( \,\frac{-2 \, \d I_{a}}{\sqrt{-h} \, \d h^{\m\n}})-    \frac{1}{\sqrt{-h}}\frac{\d I_{a}}{\d \O}    \,  \equiv 0 \quad (a= K, \L)\, .
\ee

\vskip2mm
$\bullet$ \textit{Equations of Motion}
\vskip2mm
\noindent 
To write the equations of motion following from the total action  $I_G=I_{K}+I_{\L}$ we  define\footnote{A detailed discussion of relevant conformal geometry will appear in  \cite{Dabholkar:2015b}.}
\bea\label{D-def}
E_{\m\n}[h] &:=& R_{\m\n}[h] -\half h_{\m\n} R[h]  \, ,\\
D_{\m\n}[\O, h] &:=& -(d-2) \left[ \nabla_{\m} \nabla_{\n}\,\Omega - (\nabla_{\m} \, \Omega ) \,( \nabla_{\n}\, \Omega )
  -  h_{\m\n}\left(  \nabla^{2}\Omega  + \frac{d-3}{2} |\nabla \Omega|^{2}  \right)  \right]     \, .
\eea
It is easy to check that
\be
E_{\m\n}[h] + D_{\m\n}[\O, h] =  E_{\m\n}[g] \, .
\ee
The equations of motion for the fiducial metric  give the fiducial Einstein equations: 
\bea \label{hEinstein}
E_{\m\n}[h] &=&  \frac{\kappa^{2}}{M_0^{d-2}} \left( T^{\O}_{\m\n} + T_{\m\n}^{\L} \right)
\eea
with
\be
\frac{\kappa^{2}}{M_0^{d-2}}\,T^{\O}_{\m\n} := - D_{\m\n} [\O, h]  \, \quad \textrm{and} \quad \frac{\kappa^{2}}{M_0^{d-2}} \,T_{\m\n}^{\L} =  -\L\, h_{\m\n}\,  e^{2\O}  .
\ee
The equation of motion for the $\Omega$ field is
\bea\label{omega-eom}
-2 (d-1) \nabla^{2}\, \Omega -(d-1)(d-2) | \nabla \Omega ) |^{2} + R[h]- \frac{2\,d \,\L}{d-2} e^{2 \,\Omega}&=& 0 \, .
\eea
As a consequence of the Ward identities, the equation of motion for $\O$ is  automatically satisfied if the fiducial Einstein equations are satisfied. 
In terms of the physical metric,  (\ref{hEinstein}) becomes
\be
E_{\m\n} [g]= - \L\, g_{\m \n}\, ,
\ee
which is the physical Einstein equation.
Similarly,  (\ref{omega-eom})  can be recognized as the trace of the physical Einstein equation:
\be
R[g] =  \frac{2\,d \,\L}{d-2} \, .
\ee

\subsection{Classical Cosmology}
 
 A homogeneous and isotropic universe is described by the 
Robertson-Walker metric. For a  spatially flat spacetime, one can choose a gauge in which the fiducial line-element is of the form 
\begin{equation}\label{rob-walker}
ds^{2} = -d\t^{2} + \delta_{ij} dx^{i }dx^{j}\, 
\end{equation}
on the product space $\mathbb{R} \times \mathbb{R}^{d-1}$ where  $\d_{ij}$ is the flat metric on the spatial slice and $\t$ is the conformal time.  
The physical metric has a scale factor 
\be
a(\tau) = e^{\O(\t)} \, .
\ee
The conformal time $\tau$ is related to the comoving cosmological time by
\be
d\tau = \frac{dt}{a(t)} \, .
\ee

Consider a universe filled with a perfect fluid of  energy density $\rho$ and pressure $p$.  The classical evolution of the universe is governed by the first Friedmann equation
\bea\label{Friedmann1}
H^{2} =  \frac{ 2 \kappa^{2} \r } {(d-2)(d-1) M_{0}^{d-2}}  \, 
\eea
and  conservation of  the momentum tensor implies
\be\label{Conservation}
\dot{\rho}= - (d-1)  (p + \r) H \, .
\ee
where $H:= \dot{a}/a$ as usual. If the fluid satisfies the barotropic equation of state  $p = w\r$ for some constant barotropic index $w$, then the solutions to (\ref{Friedmann1}) and (\ref{Conservation}) are given by 
\be\label{density}
\rho(t)  = \rho_{*} (\frac{a}{a_{*}})^{ -\g} \, , \qquad  a(t) = a_{*}(1 + \frac{\gamma}{2} H_{*} t) ^{\frac{2}{\g}} \, ,
\ee
where $\r_{*}$, $H_{*}$, $a_{*}$ are  the initial values of various quantities at $t=0$, and  $\g := (d-1) (1 + w)$. For the classical momentum tensor of the cosmological term we have $\r_{*} = \l M_{0}^{d}$,  $w=-1$ and $\g =0$. As  $\g \rightarrow  0$, one obtains nearly de Sitter spacetime with nearly  constant density.

 \subsection{Classical Gravity and Cosmology near Two Dimensions}
 
 Consider  the total gravitational action $I_{G}$ in the Weyl invariant formulation, which  is the sum of the Einstein-Hilbert action  \eqref{Om-action1} and the cosmological term \eqref{Om-action2}. We would like to analyze the renormalization of this action near two dimensions.
For this purpose, we first consider the classical action in $d=2+\epsilon$.  To simplify the notation  we henceforth use $R_{h}$ instead of $R[h]$ for the Ricci scalar associated with the metric $h$. Keeping only terms at most linear in $\e$ and using the rescaling \eqref{dimless} for the constants, we find
\be\label{Omega-action2}
I_{G}=\frac{M_{0}^{\e}}{2\kappa^{2}}\int d^{2 +\e}x \,  \sqrt{-h} \,\left(  R_{h}  + \epsilon\,(\,| \nabla \Omega| ^2 + R_h \Omega\,   ) + \ldots \,\right) 
-\l\, M_0^{2+\e}\, \int d^{2 +\e}x \,  \sqrt{-h}   \, e^{(2+\e)\Omega} \, .
\ee\, 
To make contact with Liouville theory in the next subsection, we define $q$ and $\mu$ by
\be 
\kappa^2=\frac{2\pi \epsilon}{q^2}, \qquad \l\, M_0^2=\m.
\ee
The action then takes the form
\be\label{total-action}
I_{G}= \frac{q^2}{4\pi}\int d^{2}x \,  \sqrt{-h} \,\left( \, \frac{R_{h}}{\e}\,+\,| \nabla \Omega|^2 + R_{h} \,\Omega  - \frac{4\pi\m}{q^2}  \, e^{2\Omega}  \right)  \, + \mathcal{O}(\e).
\ee
This action is manifestly coordinate invariant and also Weyl invariant to this order in $\e$. As a result it satisfies both Ward identities (\ref{Diff-WT}) and (\ref{Weyl-identity}). 

 Note that the cosmological evolution equations depend analytically on $d$, so one can  `dimensionally continue' them. Near two dimensions, they can be derived either by varying this action or directly by taking the limit of  (\ref{Friedmann1}) and (\ref{Conservation}) to obtain
\be
H^{2} =  \frac{ 2 \k^2 } {\e } \r_{\L} =  \frac{4\pi}{q^2}  \, \r_{\L}\, ,
\ee
and 
\be\label{Conservation2}
\dot{\rho}_{\L}= - (1+ w_{\L}) H \rho_{\L}  \, .
\ee
For the classical cosmological fluid $w_{\L}= -1$,  the energy density is constant $\r_{\L}(t) = \m$, and the Hubble scale is given by $H^2= 4\pi \m/q^2$. 
The physical metric corresponds to the de Sitter metric in cosmological coordinates with scale factor  $a(t) = a_{*}e^{Ht}$.
  
\subsection{Relation to Timelike Liouville Theory}

To compare with the Liouville action, we  define $\chi := q \,\Omega $  so that the kinetic term is  canonically normalized. The $\e$-independent part of our  action \eqref{total-action}  
then becomes
\be\label{timelike-liouville}
{I_{TL}[\chi, h]= \frac{1}{4\pi }\int d^{2}x \sqrt{-h} \,\left( | \nabla \chi|^2 + q \, R_{h}\, \chi  - 4\pi\m  \, e^{2\b \chi}  \right)} \, .
\ee
Note that the kinetic term has a `wrong sign' because in our conventions the metric has mostly positive signature. For this reason, $\chi$ is called  `timelike'  by analogy with the field corresponding to the time coordinate of target spacetime on the two-dimensional  worldsheet of a string \cite{Strominger:2003fn, Schomerus:2003vv, Zamolodchikov:2005fy, McElgin:2007ak, Harlow:2011ny,Ribault:2015sxa}. In the classical theory $\beta = 1/q$ but we keep it  as a free parameter in anticipation of quantum corrections. 

To  `analytically continue' to the usual `spacelike' Liouville theory we define
\be\label{analytic-continuation}
Q=i q \, , \qquad \vf=i \chi\, ,\qquad  b=-i \b \, ,
\ee
to obtain an action with the right-sign kinetic term:
\be\label{spacelike-liouville}
{I_{L}[\varphi, h]= -\frac{1}{4\pi }\int d^{2}x \,  \sqrt{-h} \,\left( | \nabla \vf|^2 + Q \, R_{h}\, \vf  + 4\pi\m  \, e^{2b \vf}  \right)}\, .
\ee
The Weyl transformations are given by
\be\label{chi-weyl}
h_{\m\n} \rightarrow e^{2\xi(x)}\, h_{\m\n} \quad \textrm{and}\quad  \chi \rightarrow \chi -q\,\xi(x) \quad \textrm{or}\quad  \vf \rightarrow \vf -Q\,\xi(x)  \, .
\ee
The charges are determined by requiring Weyl invariance of the first two terms of the actions (\ref{timelike-liouville}) and (\ref{spacelike-liouville}). Note though  that the these two terms  are not strictly Weyl-invariant under (\ref{chi-weyl}), but their Weyl variation is field independent with this charge assignment. Hence the equations of motion are Weyl-invariant. This is the origin of the conformal invariance of Liouville theory. Since we are  interested in the analogy to the higher-dimensional Weyl compensator, it is preferable to include  the first term of order $1/\e$ in (\ref{total-action}) so that not just the equations of motion but  the action itself is manifestly invariant under (\ref{chi-weyl}). Weyl invariance of the cosmological term then requires that classically
$q = 1/\b$ or $Q= 1/b$. 
As we discuss in \S\ref{QuantumGravity}, this relation is modified in the quantum theory because of the anomalous Weyl dimension of the cosmological constant operator. 

To discuss renormalization in the quantum theory, it is convenient to work in Euclidean space, obtained by doing a Wick rotation\footnote{To perform a Wick rotation in curved spactime, it is convenient to regard Euclidean space and Lorentzian spacetime as different real slices of a complexified spacetime. Wick rotation is then a complex coordinate transformation  $t=-i t_E$ under which all tensors transform as usual. In Lorentzian spacetime, the path integral measure is $e^{iI}$  and the spacetime measure is $\sqrt{-h}$.  In Euclidean space the path integral measure is $e^{-S}$ and the spacetime measure  is $\sqrt{h_{E}}$. Using the fact that $\sqrt{-h_{E}}=-i  \sqrt{h_E}$, we obtain $I \rightarrow -S$ with all tensors the same except $\sqrt{-h}$ replaced by $\sqrt{h_{E}}$.}. We denote the Lorentzian actions by $I$ and the Euclidean actions by $S$. The Euclidean action for timelike Liouville is 
\be\label{Euclidean Timelike Liouville action}
S_{TL}[\chi, h]=  \frac{1}{4\pi }\int d^{2}x \,  \sqrt{h} \,\left(- | \nabla \chi|^2 - q \, R_{h}\, \chi  + 4\pi\m  \, e^{2 \b \chi}  \right) \, .
\ee
For spacelike Liouville it is
\be\label{Euclidean Spacelike Liouville action}
S_{L}[\varphi, h]= \frac{1}{4\pi }\int d^{2}x \,  \sqrt{h} \,\left( | \nabla \vf|^2 + Q \, R_{h}\, \vf  + 4\pi\m  \, e^{2 b \vf}  \right) \, .
\ee 
Timelike Liouville theory as a two-dimensional model for cosmology has been considered earlier from different perspectives in \cite{Polchinski:1989fn,Cooper:1991vg, DaCunha:2003fm, Martinec:2014uva}.
  
 \section{Quantum Gravity Near Two Dimensions \label{QuantumGravity}}
 
 We work `near' two dimensions in the spirit of the $\epsilon$ expansion near four dimensions \cite{Wilson:1973jj}. Practically, this means we renormalize the theory  in two dimensions but keep $\e$ nonzero. As noted earlier, the first term of order $1/\e$ in the action (\ref{total-action})  makes the Weyl invariance manifest just as in higher dimensions.  This makes the  generalization to higher dimensions  more apparent.
 
  \subsection{Renormalization of the Cosmological Constant Operator}\label{Renorm}
 
The cosmological constant operator $e^{2\b\chi}$  is a composite operator and must be renormalized in the quantum theory. It is  most convenient to carry out this renormalization in  spacelike Liouville theory in Euclidean space.   The analytic continuation of these results to  timelike Liouville and its Lorentizan interpretation will be discussed later.  

The Liouville action $S_{L}$  contains a nonpolynomial exponential interaction. The  cosmological constant operator  should in principle be regularized in the interacting theory defined by this action. However, it is well-known  \cite{David:1988hj, Distler:1988jt,Polchinski:1990mh,Seiberg:1990eb,DHoker:1990ac,DHoker:1990md} that free-field normal ordering removes all short-distance divergences of the theory. In other words, the anomalous dimension of the cosmological constant operator in the fully interacting theory is the same  as for a much simpler theory of a free boson. Ultimately, this claim is justified by exact results obtained using the conformal bootstrap \cite{Belavin:1984vu,Dorn:1991dr, Zamolodchikov:1995aa,Teschner:2001rv, Pakman:2006hm} and agrees with the KPZ critical exponents \cite{Knizhnik:1988ak} computed using matrix models \cite{Brezin:1990rb, Douglas:1989ve,Gross:1989vs}, light-cone quantization \cite{Polyakov:1987zb}, and canonical quantization \cite{Curtright:1982gt,Braaten:1983np, Gervais:1981gs,Gervais:1982nw}.

Anomalous dimensions of exponentials of free fields have been studied extensively in string theory and two-dimensional quantum gravity \cite{Polyakov:1981rd}.  By the state-operator correspondence,  such exponentials correspond to momentum eigenstates.  To obtain the anomalous dimension, it is usually adequate to perform renormalization in flat space  by  normal ordering  \cite{Polchinski:1986qf, Polchinski:1998rq}. However, we are interested here in all three components of the quantum momentum tensor given  by metric variation of the quantum effective action. We thus require the metric dependence of the renormalized operator for an arbitrary curved metric. 

Renormalization of the cosmological constant operator in curved spacetime  has been well studied in the literature \cite{Polyakov:1981rd,Onofri:1982zk,Polchinski:1986qf, DHoker:1987bh,DHoker:1988aa}. Since it is of crucial importance for our conclusions, we present below a somewhat lengthy derivation taking into account some of the subtleties both in the UV and in the IR. New conceptual questions of  interpretation arise  in  continuing the Euclidean computations to Lorentzian spacetime which we discuss in \S\ref{Lorentzian}. We then write down the quantum effective action for this renormalized term, compute the quantum  momentum tensor and check explicitly that the Ward identities are satisfied\footnote{In \S\ref{Alternative} we  reverse the logic and compute the stress tensor from the anomalous trace using the Ward identities.}. 

Consider the correlation function of exponentials in a free theory  in a curved background:
\be\label{n-point}
\cA^{0}(x_{1}, \ldots, x_{n}) := \langle \prod_{i=1}^{n} e^{2 a_{i} \vf(x_{i})} \rangle = \int D\vf \,e^{-S[\vf, h]} \prod_{i=1}^{n} e^{2 a_{i} \vf(x_{i})} = \int D\vf \,e^{-S[\vf, h] + \int d^{2}x \sqrt{h} \,J(x)\, \vf(x) } \, .
\ee 
The superscript `$0$' is a reminder that this is a bare correlation function with the action
\be
 S[\vf, h]=\frac{1}{4\pi} \int d^{2} x \sqrt{h}  \, |\nabla \vf|^2\, , \quad \textrm{and} \quad J(x) = 2\sum_{i=1}^n a_{i}\, \d^{(2)} (x,  x_{i}).
\ee
We have set $Q=0$ in (\ref{Euclidean Spacelike Liouville action})  so that the Liouville field is neutral under (\ref{chi-weyl}). While the classical dimension depends on $Q$, the \textit{anomalous} dimension of our interest is independent of $Q$. Using Wick's theorem one obtains
\be\label{n-point-answer-bare}
\cA^{0} (x_{1}, \ldots, x_{n}) = \exp{\left[4\pi \sum_{i, j}a_{i} a_{j}\,G_{h}(x_{i}, x_{j})\right]}
\ee
where $G_h$ is the scalar Green function\footnote{We define the  Laplacian  as $-\nabla_{h}^{2}$  so that it is a positive operator.}:
\be\label{green}
-\nabla_h^2 G_{h}(x,y) =\d^{(2)}_{h}(x,y)=\frac{\d^{(2)}(x-y)}{\sqrt{h}}\, .
\ee 
In general, the Green function for an arbitrary metric $h_{\m\n}$ is hard to compute. However, in  two dimensions,  $\nabla_{h}^2= e^{-2\Sigma_{h}} \nabla_{\d}^2$, and hence the (noncompact) Green equation is Weyl invariant. Moreover, in two dimensions one can always choose a conformal coordinate frame so that
\be\label{conformal-frame}
h_{\m\n} = e^{2\Sigma_{h}} \d_{\m\n}\, .
\ee  
The Green function is then given by the flat space Green function. The latter is known to be  infrared divergent\footnote{On  a compact manifold  there is no need for an IR regulator but the Laplacian has a zero mode which has to be treated  carefully \cite{Onofri:1982zk, DHoker:1988aa}.}. To regulate this divergence, consider the class of asymptotically flat metrics so that $\Sigma_{h}(x) \rightarrow 0$  as  $|x| \rightarrow \infty$. Introduce an IR cutoff  by restricting $\mathbb{R}^{2}$ to a disk in the flat metric
\be\label{ir-cutoff}
|x|^{2}:= \d_{\m\n}x^{\m}x^{\n} \leq  R^{2} :=1/m^{2}
\ee 
and impose Dirichlet boundary conditions at $|x| = R$. The resulting Green function is 
\be\label{noncoincidentGreen}
G_h(x,y) = G_\d(x,y)= -\frac{1}{4\pi}\ln{(m^{2}|x-y|^2)} \,  \quad \textrm{for} \quad x \neq y \, ,
\ee 
where we have ignored the contribution from image charges which are negligible in the limit  $R\rightarrow \infty$. The boundary condition and the Green equation are both invariant under Weyl transformations that asymptote to unity for $|x| \rightarrow \infty$. For all metrics related by such Weyl transformations, the Green function is the same as above. The Green function is invariant also under constant Weyl transformations if we  scale the IR cut-off  at the same time.

\begin{figure}[htbp] 
\begin{center}
{\includegraphics[scale=1.50]{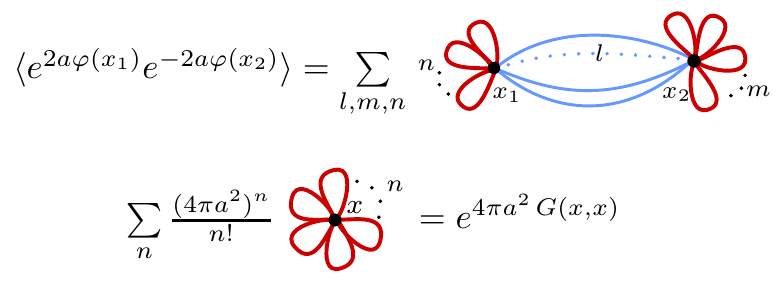}}
\caption{\textit{The red daisies at each point come from self-contractions.  Each petal of a daisy  is a  coincident Green function  and the sum over these daisies  gives a divergent exponential.
}}
\label{daisy}
\end{center}
\end{figure}

Naively, the Weyl invariance of the Green function implies that the IR-regulated $n$-point function is Weyl invariant. But on general grounds, one expects that regularization of UV divergences will 
 introduce a dependence on the  metric that can violate the Weyl symmetry.   To compute this anomalous Weyl variation, we rewrite  the $n$-point function  as
 \be\label{n-point-answer}
\cA^{0}(x_{1}, \ldots, x_{n}) = \prod_{i }e^{4\pi a_{i}^{2}\, G^{0}_{h}(x_{i}, x_{i})} \cdot  \exp{\left[4\pi\sum_{i \neq j}a_{i} a_{j}\,G_{h}(x_{i}, x_{j})\right]} \, .
\ee
 As it stands, this is only a formal expression that is  not well-defined.
The prefactor is a product over exponentials of Green functions evaluated at the same points, which are divergent.  We have  therefore added a superscript to underscore the fact that the coincident Green functions are bare quantities.  The origin of the UV divergence is clear from (\ref{n-point-answer}): each exponential is a composite operator involving  products of the fundamental scalar field with divergent self-contractions. This is shown diagrammatically in Fig.\ref{daisy} for a two-point function.

To regulate this divergence we rewrite the coincident Green function as 
\be\label{hk-reg-green}
G_h^\ve(x,x)  =  \int d^2y \,\sqrt{h}\,  \d_h^{(2)}(x, y) \, G_h(y, x)
= \int d^2y \,\sqrt{h}\, K_h(x, y;\ve) \, G_h(y, x)\, ,
\ee
where we have replaced the delta function by the heat kernel with a short time cutoff\footnote{We use $\ve$ for the short-time cutoff and $\e$ for the dimensional regulator.}  $\varepsilon$ since
\be
K_h(x,y;\ve) \rightarrow \d_h^{(2)}(x,y) \, \quad \textrm{as} \quad \ve \rightarrow 0\, .
\ee
The short-time expansion of the heat kernel can be obtained using standard methods in terms of Seely-deWitt coefficients. The computations are simpler  in the conformal frame (\ref{conformal-frame}). The leading behavior is given by
\be\label{2d-leading-hk}
K_{h}(x, y;\ve) = \frac{1}{4\pi \ve}{\exp{\left[\frac{e^{2\Sigma(x)}|x-y|^{2}}{4\ve }\right]}}  \left(1 + \ldots \right) \, \, .
\ee
The regularization separates the two points by a distance of order $\sqrt{\ve}$. 
 Using this expansion one obtains 
\be\label{green-reg}
G_{h}^{\ve}(x, x) =\frac{1}{2\pi}\Sigma_h(x)-\frac{1}{4\pi} \ln (4\, e^{-\g}\, m^{2}\ve)\, ,
\ee
where $\g$ is the Euler-Mascheroni constant. 
More details and an alternative derivation in dimensional regularization are given  in \S\ref{appendixA}. 

The final expression (\ref{green-reg}) for the regulated coincident Green function is not  yet completely coordinate invariant because  $\Sigma_{h}(x)$ is the conformal factor of the metric only in the conformal frame. One might be tempted to write $\Sigma_{h}(x)$  in terms of  the determinant of the metric, but this cannot be correct because $G_{h}^{\ve}(x, x)$  must be a scalar, whereas the determinant of the metric is a scalar density. To obtain a  manifestly coordinate-invariant scalar expression  we note that
\be
R_{h}= e^{-2\Sigma_{h}(x)}\left( R_{\d}-2\,\nabla_\d^2 \Sigma_{h}(x)\right)=-2 \,\nabla_h^2 \Sigma_{h}(x)\, .
 \ee
By solving this Poisson equation, the conformal factor can be written as
\be\label{sigma-nonlocal}
\Sigma_{h}(x)=\half \int d^2y \,\sqrt{h}\, G_h(x,y)\, R_h(y)\, .
\ee
A manifestly coordinate-invariant and  regularized coincident Green function is then
\be
G_{h}^{\ve}(x, x) =\frac{1}{4\pi}\int d^2y \,\sqrt{h}\, G_h(x,y)\, R_h(y)-\frac{1}{4\pi} \ln (4\, e^{-\g}  m^{2}\, \ve)\, .
\ee
Renormalization now consists in simply adding $ \frac{1}{4\pi} \ln (4\, e^{-\g} M^{2}\, \ve)$ so the divergent term with $\ve$ is removed. Since $\ve$ is a pure number\footnote{It is convenient to regard all quantities including spacetime coordinates and mass scales like $m$ as dimensionless,  measured in units of the fundamental UV scale $M_{0}$ introduced earlier, which we can set to one.} independent of coordinates and the metric, this procedure is manifestly coordinate invariant and \textit{local}. Renormalization has introduced an arbitrary scale $M$. The renormalized coincident Green function is then given by
\be\label{renormalized-green}
G_{h}(x, x) =\frac{1}{4\pi}\int d^2y \,\sqrt{h}\, G_h(x,y)\, R_h(y) + \frac{1}{4\pi} \ln (M^{2}/m^{2})\, 
\ee
which however is not a local functional of the metric\footnote{This nonlocality was emphasized earlier in \cite{Banks:1989qe}.}. 
The renormalized  $n$-point function  can now be obtained simply by replacing 
the bare coincident Green function $G_{h}^{0}(x, x) $ in (\ref{n-point-answer}) by the renormalized coincident Green function (\ref{renormalized-green}). The resulting answer is finite and independent of $\ve$, but one which depends on the renormalization scale.   It corresponds to a multiplicative renormalization of each of the bare exponentials:
\be
[e^{2a \vf (x)}]^{\ve}_{h} := e^{{-a^{2}\ln{ (4\, e^{-\g} M^{2}\ve)} }}[e^{2a \vf(x)}]_{h} := Z_{a}(M) \, [e^{2a \vf(x)}]_{h}
\ee
where the notation $[\cO]_{h}^{\ve}$ indicates an operator $\cO$ regularized using the metric $h$ and cutoff $\ve$, whereas  $[\cO]_{h}$ without superscript indicates the renormalized version of the same operator.   We have defined the multiplicative operator renormalization  $Z_{a}(M)$ to make contact with the usual flat-space renormalization. 
Even though this procedure is manifestly local  and coordinate invariant, it is not Weyl-invariant because it depends on the background metric. With this renormalization prescription, the  $n$-point function renormalized using the $h$ metric is given by
\bea\label{n-point-answer2}
\cA_{h}(x_{1}, \ldots, x_{n}) = m^{-2\,(\sum_{i} a_{i})^{2}} \prod_{i } (Me^{\Sigma_{h}(x_{i})})^{2a_{i}^{2}}\exp{\left[- \sum_{i \neq j}a_{i} a_{j}\log |x_{i}- x_{j}|^{2}\right]} \,.
\eea
The first factor simply imposes momentum conservation: the correlation function vanishes unless the total momentum is zero\footnote{Operators with positive Weyl weight are defined only for $a_{i}= i k_{i}$ for real $k_{i}$. They correspond to normalizable charge eigenstates in the Hilbert space. The prefactor is then a positive power of $m$ which vanishes as $m\rightarrow 0$. For operators with negative weight the correlation functions diverge at large separation. The corresponding states are not normalizable and have to be interpreted using an analog of the  Gelfand triple \cite{Teschner:2001rv}.}. This is to be expected because  momentum is the charge corresponding to a continuous global symmetry $\vf \rightarrow \vf + c$ which cannot be  spontaneously broken in two dimensions by the Coleman-Mermin-Wagner theorem. Imposing momentum conservation, the final expression for the renormalized $n$-point function is given by
\bea\label{n-point-answer3}
\cA_{h}(x_{1}, \ldots, x_{n}) &=&  \prod_{i } e^{2 a_{i}^{2}\Sigma_{h}(x_{i})}
\prod_{i \neq j} \frac{1}{\left(M|x_{i} - x_{j}|\right)^{2a_{i}a_{j}}}\, .
\eea
For $\Sigma_h(x)= -\log(M)$, we obtain the familiar answer from flat space conformal field theory.

The $n$-point correlators renormalized in two different metrics are related by
\be\label{n-point-compare}
\cA_{h'}(x_{1}, \ldots, x_{n}) = \prod_{i } 
e^{2 a_{i}^{2}\left(\Sigma_{h'}(x_{i}) - \Sigma_{h}(x_{i})\right)}
\cA_{h}(x_{1}, \ldots, x_{n}) \, . 
\ee
This follows  from  (\ref{n-point-answer3}) and the fact that the non-coincident Green function given by (\ref{noncoincidentGreen}) is independent of the metric.
Interpreting the correlation function in operator language, we conclude that the exponential operator renormalized using the metric $h'$ is related to the one renormalized using the metric $h$ by 
\be\label{op-compare}
[e^{2a \hat \vf(x)}]_{h'}= 
e^{2 a^{2}\left(\Sigma_{h'}(x) - \Sigma_{h}(x)\right)}\, 
[e^{2a \hat \vf(x)}]_{h}\, 
\ee
where the hatted variable denotes a quantum operator rather than a classical field. 

The cosmological constant operator in Liouville theory corresponds to $a=b$. 
The Weyl transformation of the renormalized cosmological constant operator has an \textit{anomalous} contribution from  (\ref{op-compare}) as computed above because of the implicit dependence on the metric through renormalization.  In addition, for nonzero $Q$, there is also a \textit{classical} contribution because of the explicit dependence on $\vf$ which transforms as in (\ref{chi-weyl}). The net Weyl transformation is
\be\label{weyl-transform-renorm2}
[e^{2b \hat \vf(x)}]_{h} \rightarrow e^{-(2bQ - 2b^2 )\, \xi(x)} \, [e^{2 b \hat \vf(x)}]_{h}\, .
\ee
We interpret  $2 b Q$ as the classical Weyl weight and $-2b^{2}$ as the anomalous Weyl weight. 

\subsection{Lorentzian Interpretation \label{Lorentzian}}

At a formal level, analytic continuation to timelike Liouville in  Lorentzian spacetime  is straightforward using (\ref{analytic-continuation}) and a Wick rotation.  We will use the same covariant expression for $\Sigma_{h}$:
\be\label{lor-sigma}
\Sigma_{h}(x) = \half \int d^2y \,\sqrt{-h}\, G_h(x,y)\, R_h(y)
\ee
where the Green function\footnote{A Wick rotation would give a factor of $i$ for the measure and a factor of $-i$ for the Green function. In (\ref{lor-sigma}) and (\ref{lor-green}) we drop both factors.}  is the solution of the Lorentzian Green equation without any $i$:
\be\label{lor-green}
-\nabla^2 G_{h}(x,y) =\d^{(2)}_{h}(x,y)=\frac{\d^{(2)}(x-y)}{\sqrt{-h}}\, .
\ee
 
Physical interpretation in the Lorentzian  signature is subtle. We discuss below some of the puzzles that one encounters in interpreting the  Lorentzian action and their resolutions.
\begin{myitemize}

\item \textit{Choice of the Green function:} The Lorentzian Green function appearing in the expression \eqref{lor-sigma} for the $\Sigma_{h}$  depends on the choice of the boundary condition. The Euclidean Green function in  \S\ref{QuantumGravity} is  unique and usually it would continue to the Feynman propagator under a Wick rotation.
However, one could equally well choose retarded or advanced boundary conditions, which would lead to very different physics.  Which of these Green functions is physically relevant? We are eventually interested in using the  quantum effective action to study classical evolution equations. Appearance of Feynman propagators in the effective action would lead to  non-causal dynamics because it would involve negative energy modes traveling backward in time. Such an effective action would be unphysical. However, in  time-dependent situations as in cosmology, the in-vacuum and the out-vacuum are in general different. A natural object to consider  is not the usual in-out effective action,  but the in-in effective action in the Schwinger-Keldysh formalism \cite{Schwinger:1960qe,Keldysh:1964ud}. It is known that one can obtain the in-in effective action from the in-out one by replacing   Feynman propagators by  retarded Green functions \cite{Calzetta:1986ey,Jordan:1986ug, Barvinsky:1987uw,Higuchi:2010xt}.

\item \textit{Choice of the vacuum:} In canonical formalism in the Lorentzian theory, the choice of the metric used for renormalization corresponds to the choice of the vacuum, as we discuss below. We choose  the Minkowski metric $\eta_{\m\n}$ as a reference metric, which corresponds to  $\delta_{\m\n}$ under Euclidean continuation.  
Continuation of (\ref{op-compare}) gives  the following equation for  the renormalized cosmological constant operator in Lorentzian spacetime
\be\label{renormalized-cosmo-op}
[e^{2\b \hat \chi(x)}]_h = e^{-2\b^2\, \Sigma_{h}(x)} \, [e^{2\b \hat \chi(x)}]_{\eta}\, .
\ee

As it stands, (\ref{renormalized-cosmo-op}) is an operator equation with a quantum operator $\hat \chi(x)$ in the exponent. In  the cosmological term in the quantum effective action, we would like to regard $ \chi(x)$ as a classical field. This is achieved using the background field method by replacing $\hat \chi(x)$ by $\chi(x) + \hat\chi_{q}(x)$. The unhatted variable is a classical background field and the hatted variable is the fluctuating quantum field.  For the free action, $\hat\chi_{q}(x)$ is also a free field\footnote{In the background field method one chooses an external source as a functional of the background field in such a way as to cancel all tadpoles. See \cite{Abbott:1981ke, Jackiw:1974cv} for a concise summary.}. We then have the relation
\be\label{cosmo-field1}
[e^{2\b \,\left(\chi(x) +\hat\chi_{q}(x)\right)}]_h = e^{2\b \chi(x) -2\b^2\, \Sigma_{h}(x)} 
\, [e^{2\b\hat \chi_{q}(x)}]_{\eta}\, .
\ee
The passage to the in-in effective action still requires a choice of the in-vacuum to compute  the in-in  matrix element of this operator.  Which state should one choose as the  in-vacuum? The choice of the vacuum is a deep and unresolved question in cosmology since it concerns the initial state in which the universe `got prepared'. Even in a free theory, there are many possible Fock vacua that are \textit{a priori} equally valid as initial states. In general, the Fock vacuum  depends on the metric used to define the Klein-Gordon inner product. This inner product is essential to obtain  the division of  the modes of the Klein-Gordon operator (on a globally hyperbolic spacetime) into positive-frequency and negative-frequency modes, and hence to determine the class of annihilation operators that should annihilate the vacuum. A conventional choice is the `Bunch-Davies'  vacuum $|\eta\rangle$, obtained using the Klein-Gordon inner product defined with respect to the flat Minkowski metric $\eta$. This would coincide with the conformal or the adiabatic vacuum \cite{Birrell:1982ix}. 

\end{myitemize}
In summary, a physically reasonable interpretation of the Lorentzian continuation requires that we consider  the in-in quantum effective action  and hence use retarded Green functions. The cosmological term can be regarded as the expectation value in the $\eta$-vacuum of the operator renormalized using the $h_{\m\n}$ metric. We denote this \textit{classical} quantity by $\cO^{\b}_{h}$:
\be\label{cosmo-field}
\cO^{\b}_{h} := \langle \eta | [e^{2\b (\chi(x) +\hat\chi_{q}(x))}]_h | \eta \rangle = e^{2\b \chi(x) -2\b^2\, \Sigma_{h}(x)} \langle \eta | [e^{2\b\hat \chi_{q}(x)}]_{\eta} | \eta \rangle = e^{2\b \chi(x) -2\b^2\, \Sigma_{h}(x)}
\, ,
\ee
where in the second equality  we have used  the fact that, in the Hamiltonian formalism,   renormalization in the metric $\eta$ corresponds to normal ordering with respect to the $\eta$-vacuum, and hence the expectation value of the exponential equals one. 

With these ingredients, the integrated renormalized cosmological term in the quantum effective action  takes the final form
\be\label{renormalized-integrated-cosmo-op1}
I_\L[\chi, h] =-\m\, \int d^{2}x \, \sqrt{-h} \, \cO^\b_{h}  \,.
\ee
The Weyl transformation of $\cO^{\b}_{h}$ is given by
\be\label{weyl-transform-renorm}
\cO^{\b}_{h} \rightarrow e^{-(2\b q + 2\b^2 )\, \xi(x)} \, \cO^{\b}_{h}\, .
\ee
Since the integration measure  $\sqrt{-h}$ has Weyl weight $-2$, quantum Weyl invariance of the integrated cosmological term implies 
\be\label{relation-qb}
2\b q + 2\b^{2} -2 =0  \, \quad \textrm{or} \quad q=\frac{1}{\b}-\b.
\ee
In Liouville literature,  $2\b^{2}$ is sometimes referred to as the `\textit{anomalous gravitational dressing}' of the identity operator.
Recall that classically, Weyl invariance required that $\beta = 1/q$. We can regard $1/q$ as the coupling constant and interpret our results as quantum corrections to $\b$ so that Weyl invariance is maintained at the quantum level:
\be\label{b-q}
\b=\frac{q}{2}\left( -1+\sqrt{1+\frac{4}{q^2}}\right)=\frac{1}{q}-\frac{1}{q^3}+ \frac{2}{q^5} + \ldots \, .
\ee

\subsection{Nonlocal Quantum Effective Action}

With this  interpretation, the cosmological term in Lorentzian spacetime in  terms of  $\Omega$ becomes
\be\label{renormalized-integrated-cosmo-op}
I_\L [\O, h] =  -\m\,\int d^{2}x \, \sqrt{-h}\, e^{-2\b^2\, \Sigma_{h}}\,e^{2\,\b\,q\,\O} \, = -\m\,\int d^{2}x \, \sqrt{-h}\, e^{-2\b^2\, (\O +\Sigma_{h}) }\,e^{2\,\O} \,.
\ee
 The total gravitational effective action for the background fields is given by
\be\label{qaction1}
I_{eff}[\O, h]=\frac{q^2}{4\pi}\int d^{2}x \sqrt{-h} \,\left( \frac{R_{h}}{\e}+ |\nabla\O |^2+R_{h}\,\O  -\frac{4\pi\m}{q^{2}} \,  \,e^{2\,\O}\, e^{-2\b^2\,( \O+\Sigma_{h})} \right)\, .
\ee
One can choose  a gauge in which $\Omega =0$ so that the fiducial metric can be identified directly with the physical metric defined by \eqref{metric-relation}. In this physical gauge, the action takes the form
\be\label{qaction2}
I_{eff}[g] =\frac{q^2}{4\pi}\int d^{2}x \sqrt{-g} \,\left( \frac{R_{g}}{\e}  -\frac{4\pi\m}{q^{2}} \, e^{-2\b^2\, \Sigma_{g}} \right)\, .
\ee

The effective action is nonlocal  and one might  worry about possible ghosts. In fact, in the local formulation  described in \S\ref{Local2}, one of the auxiliary fields has a negative kinetic term. Quantization of this degree of freedom would typically lead to a violation of  both causality and unitarity. The correct point of view is to regard the quantum effective action as the {result} of having evaluated a path integral in the presence of a \textit{classical} background field. Thus, this effective action is not to be quantized further but rather to be used to study effective dynamics classically. After imposing appropriate initial conditions, one expects a ghost-free causal evolution because the original path integral is well-defined.

In addition to the anomalous Weyl dimension of the  cosmological constant operator, there is the usual source of Weyl anomaly which is proportional to the central charge. We consider gravity coupled to conformal matter with central charge $c_{m}$.  The  Fadeev-Popov ghosts have central charge $-26$ and the Liouville field has  central charge $1$. Quantum Weyl invariance then requires that the total anomalous  central charge  vanishes:
\be
c_{total} = c_{m} -26
 + 1=0
\ee
Note that  when the first term of  order $1/\e$ term in \eqref{qaction1} is included, the classical Liouville action is \textit{invariant} under Weyl transformations. As a result, the central charge of the Liouville field is $1$ and not $(1-6q^{2})$, and  $q^{2}$ is a free parameter independent of $c_{m}$ even after imposing quantum Weyl invariance. The semiclassical limit corresponds to large $q$ independent of $c_{m}$.  This is more natural from the point of view of dimensional continuation to  $d$ dimensions.

 \subsection{Quantum Momentum Tensor}
 
The quantum momentum tensor associated with the effective action \eqref{renormalized-integrated-cosmo-op} is given by\footnote{In two dimensions, the momentum tensor obtained by varying the fiducial metric $h_{\m\n}$ for fixed $\O$ is the same as the momentum tensor obtained by varying the physical metric $g_{\m\n}$.}
 \bea
 T^\L_{\m\n}(x) &:=& \frac{-2}{\sqrt{-h}}\frac{\delta I_\L}{\delta h^{\m\n}(x)}\\
 &=&
 -\m \,\cO^\b_h(x) \, h_{\m\n} -4\m\,\b^2 \int dy  \, \Sigma_{\m\n}(x,y) \,\cO^\b_h(y) \, . 
 \eea
 The second term is the variation of  the nonlocal term:
 \be\label{general-variation-sigma}
 \Sigma_{\m\n}(x,y) :=  \frac {1}{\sqrt{-h}}\frac{\d\Sigma(y) }{\d h^{\m\n}(x)}\,=\frac {1}{\sqrt{-h}}\frac{\d\quad }{\d h^{\m\n}(x)} \,\half \int dz\, G_{yz}\,R_{h}(z)\, 
 \ee
  where $G_{xy}$ is a shorthand for $G_{h}(x, y)$ and $dy$ is a shorthand for $d^{2}y \sqrt{-h}$.
 Using the variation of the Green function computed in  appendix \S{\ref{GreenVariation}, we obtain
 \bea\label{general-variation}
2 \Sigma_{\m\n}(x, y) =  h_{\m\n} \nabla^2 G_{xy} -  \nabla_\m\nabla_\n G_{xy} -\int dz \,R_{h}(z) \left(
\nabla_{(\m} G_{yx}\,\nabla_{\n)} G_{xz} - \half h_{\m\n}\,\nabla_\a G_{yx}\,\nabla^\a G_{xz}\right)\, \nonumber
 \eea
where all derivatives and unspecified arguments of fields such as $h_{\m\n}$ correspond to the variable $x$, and we have used the fact that the Einstein tensor in two dimensions vanishes.   The final expression for the quantum momentum tensor can be written as 
\bea\label{quantum-EMtensor}
{T}^{\L}_{\m\n}(x)=-\m \, (1 -\beta^{2}) \,  h_{\m\n} \, \cO^{\b}_h(x) + 2\, \m \, \b^2 \,  S_{\m\n}(x)
\eea
where $S_{\m\n}$ is  nonlocal and traceless:
\bea\label{Stensor}
S_{\m\n}(x) &=& \int dy \, \Big[ \nabla_\m \nabla_\n- \half h_{\m\n} \, \nabla^2 \Big] G_{xy}\, \cO^\b_h(y )  \\
&+&  \int dy \, dz
\Big[\nabla_{(\m} G_{yx} \nabla_{\n)} G_{xz} - \half h_{\m\n}\, h^{\a\b}\,  \nabla_{\a}\, G_{yx}\, \nabla_{\b} \, G_{xz} \Big] \, \cO^\b_h(y )\,R_h(z) \, . \nonumber
\eea

 \subsection{Quantum Ward Identities}

We first check the Ward identity \eqref{Weyl-identity} for Weyl invariance for the renormalized cosmological term \eqref{renormalized-integrated-cosmo-op}. 
The left hand side of \eqref{Weyl-identity} evaluates to
\begin{align}
h^{\m\n} \,{T}^{\L}_{\m\n} - \frac{1}{\sqrt{-h}}\frac{\d I_{\L}}{\d \O}
=  -2\m\,(1-\b^2) \cO^\b_h + 2\,\m\,\b \,q\, \cO^\b_h  \, .
\end{align}
It vanishes precisely when $\beta$ is related to $q$ by \eqref{relation-qb}. 
This is to be expected because the Weyl Ward identity is simply the infinitesimal version of invariance under finite Weyl transformations  which is  what was used to obtain \eqref{relation-qb}. 
The important point is that unless we modify $\b$ as in \eqref{b-q} away from its classical value, the full quantum theory would be anomalous. Anomalies in Weyl invariance are unavoidable because of the  necessity to regularize the path integral. In the present context, we manage to maintain Weyl invariance at the quantum level by starting with a value of $\beta$ such that the  theory  is \textit{not} Weyl invariant at the classical level  but it \textit{becomes} Weyl-invariant at the quantum level once the anomalous variations are taken into account.

For diffeomorphisms, we do not expect any anomalies because the renormalization procedure is manifestly coordinate invariant. To explicitly check the Ward identity  we compute 
the covariant derivative of \eqref{quantum-EMtensor}. Using the commutator  in two dimensions
\be
[\nabla_\m,\nabla_\n] V^\m= R_{\m\n}V^\m= \half R_{h}\, V_\n 
\ee
and the Green equation to cancel terms, we obtain 
\bea
\nabla^\m \,T^\L_{\m\n} = -\m\,\nabla_\n \cO^\b_h-\m\,\b^2\, \cO^\b_h(x)\,\int d z\,R_h(z)\, \nabla_\n G_{xz} 
= - 2\m\,\b \,q\, \cO^\b_h\,\nabla_\n \O \, 
\eea
where in the last step we have integrated by parts and used the expression for the renormalized operator \eqref{cosmo-field}. 
This coincides with the $\O$ variation of the action
\be
\frac{1}{\sqrt{-h}} \frac{\d I_\L}{\d \O}\, \nabla_\n \O=
\left( -2\m\,\b \,q\,\cO^\b_h\right)\,\nabla_\n \O= -2\m\,\b \,q \,\cO^\b_h\,\nabla_\n \O  \, .
\ee

\subsection{Quantum Momentum Tensor from the Weyl Anomaly \label{Alternative}}

One can  derive the momentum tensor   directly using the Weyl anomaly by reversing the logic of the previous subsection.  We \textit{assume} the diffeomorphism Ward identities rather than \textit{verify} them. Our assumption is justified by the fact that our renormalization scheme used for computing the anomalous Weyl dimension is manifestly coordinate invariant and hence there is no possibility of diffeomorphism anomalies. 
The advantage of this method is that one can avoid  the intermediate step of deducing the quantum effective action  and directly obtain the quantum momentum tensor required in the equations of motion.

For this purpose it is convenient to use  the conformal frame (\ref{conformal-frame}) with lightcone coordinates\footnote{We use the  (+++) conventions of Misner, Thorne, and Wheeler. Our lightcone coordinates are $x^{\pm} := t\pm x$. The flat metric in these coordinates is $\eta_{+-}=-\half$ with $\sqrt{-\eta}=\frac{1}{2}$ and $\nabla_{\eta}^2=-4 \,\partial_+\partial_-$.}. The only nonvanishing Christoffel symbols are
\bea\label{christoffel}
\G^{+}_{++} = 2\, \partial_{+}\Sigma_{h} \, , \quad \G^{-}_{--} = 2\, \partial_{-}\Sigma_{h} \, .
\eea
In two dimensions, the momentum tensor  has only three independent components.   Diffeomorphism Ward identities  \eqref{Diff-WT} give two  equations. From the Weyl anomaly one obtains 
\be\label{T+- component}
T_{+-}=\half \,\m \, (1-\b^2)\, e^{2\Sigma_{h}}\, \cO^\b_h(x) \, .
\ee
Together we obtain three equations for all three unknowns. The diffeomorphism Ward identity for the $\n=+$ component gives
\be
 \partial_- T_{++} +\partial_+ T_{-+}-2 \,\partial_+\Sigma_{h}\, \,T_{-+}= \m \,(1-\b^2)\,\partial_+ \O\, e^{2\Sigma_{h}} \, \cO^\b_h
 \ee
 which after use of \eqref{T+- component} becomes
 \be
\partial_- T_{++} = \half \m\, \b^2 \, \partial_+  \left( e^{2\Sigma_{h}} \, \cO^\b_h \right)\, . \nn
\ee
Taking a derivative with respect to $+$ on both sides we obtain
\begin{gather}
- \nabla_{h}^2 T_{++}= 2\m\,\b^2 e^{-2\Sigma_{h}} \,\partial^2_{+} \left( e^{2\Sigma_{h}} \, \cO^\b_h \right)  \, 
\end{gather}
where $-\nabla_{h}^2 $ is the scalar Laplacian. Solving this Poisson equation we obtain
\be
T_{++}(x)= 2\m\,\b^2  \int dy \,  \partial_+^2 G_{xy} \, \cO^\b_h(y ) \, .
\ee
We rewrite the partial derivatives as covariant ones and  use the expression \eqref{sigma-nonlocal} for the $\Sigma_{h}$ factors in  the Christoffel symbols \eqref{christoffel}  to obtain a covariant expression
\be
T_{++}(x)= 2\m\,\b^2 \, \int dy \,  \nabla_+\nabla_+ G_{xy}\, \cO^\b_h(y ) \,
+\, 2\m\,\b^2 \,\int dy\,  dz \,   \nabla_+ G_{yx } \nabla_+ G_{xz}\, \cO^\b_h(y )\, R_h(z )\, 
\ee
in agreement with \eqref{quantum-EMtensor}. The component $T_{--}$ can be computed similarly. 

\section{Quantum Cosmology Near Two Dimensions \label{QuantumCosmo}}

In this section we  examine the cosmological consequences of the quantum anomalies summarized by this effective action \eqref{qaction1} 
assuming positive cosmological constant $\m  > 0$. 
 \subsection{Quantum Evolution Equations for Cosmology}

The quantum  momentum tensor appearing on the right hand side of Einstein equation is nonlocal. 
The quantum evolution equations for cosmology are no longer given by the usual Friedmann-Lema\^{i}tre equations but rather by a set of integro-differential equations. 

Surprisingly, the quantum momentum tensor simplifies considerably for the isotropic and homogeneous universes when the fiducial metric is flat and $\Omega$ depends only on time.  The retarded Green's function of the Laplacian in two-dimensional flat spacetime is given by
\be
G_{ret}(x,y )=\half \Theta(\t_{x}-\t_y-|r_{x}-r_y|)\, .
\ee
For flat fiducial metrics, the second term in  the expression (\ref{Stensor}) for  $S_{\m\n}$ vanishes and  the first term gives $S_{\m\n}=\left(\d_\m^\t\d_\n^\t +\half \eta_{\m\n}\right)\,\cO^{\b}_h(\t)$. 
The total momentum tensor is given by
\be
 T^{\L}_{\m\n}(\t)
  =-\m \left( (1-2\b^2)\,\eta_{\m\n} -2\b^2 \,\d^\t_\m \d^\t_\n\right) \cO^{\b}_h(\t)\, .
\ee
From the components of $T^\L_{\m\n}$ we can identify its density and pressure as
\bea\label{density pressure quantum cosmo}
\r_\L(\t) = \m\, e^{-2\b^2\O(\t)} \, , \quad 
p_\L(\t) = -\m\, (1-2\b^2) \, e^{-2\b^2\O(\t)} 
\eea
which imply the equation of state
\be
 p_{\L} = w_{\L} \rho_{\L} \quad \textrm{with} \quad  w_{\L} =-1+2\b^2\, .
\ee
Thus, in the semiclassical limit of small $\beta$, the  barotropic index is slightly bigger than $-1$.

Remarkably, the nonlocal quantum cosmological momentum  tensor  has reduced  to a local one with a particularly simple form corresponding to a barotropic perfect fluid. The net effect of the nonlocal quantum contribution to the momentum tensor is simply to modify the barotropic index from $-1$ to $-1+ 2\b^{2}$. 
With this simplification for the momentum tensor, the complicated integro-differential equations for cosmological evolution  reduce to the familiar Friedmann equations for a perfect fluid, but with an unusual barotropic index. Applying the formulae from our discussion of classical cosmology, in particular \eqref{density}, we see that $\g = 2\b^{2}$ for the vacuum fluid. We arrive at the conclusion that the quantum cosmological term leads to  an expanding universe with  decaying vacuum energy density and power law expansion 
\be
\rho(t)  = \rho_{*} (\frac{a}{a_{*}})^{ -2\b^{2}} \, , \qquad a(t) = a_{*}( 1+ \b^{2} H_{*} t)^{\frac{1}{\b^{2}}} \, .
\ee
\subsection{Cosmological Implications of the Quantum Decay of Vacuum Energy}

These theoretical conclusions have potentially far-reaching implications for addressing some of the fundamental puzzles in modern cosmology \cite{Dabholkar:2015qhk,Dabholkar:2015b}. We briefly comment on some of these consequences that can generalize to higher dimensions in a model-independent way. Analogous analysis of four-dimensional quantum cosmology will be presented in \cite{Dabholkar:2015b,Bautista:2015b, Dabholkar:2015}.

The  quantum decay of vacuum energy can provide a dynamical solution to the cosmological constant problem \cite{Weinberg:1988cp,Straumann:1999ia,Weinberg:2000yb,Witten:2000zk,Polchinski:2008ux}. 
One can imagine that the universe starts off with a very large cosmological constant. The initial magnitude $\rho_{*}$ of the vacuum energy density is of the order of $M_{0}^{2}$ for the cutoff scale $M_{0}$ which can be of the order of the string scale or the scale of supersymmetry breaking. Classically, one would obtain  exactly de Sitter  spacetime with exponential expansion and constant energy density. With even a very small value of the anomalous gravitational dressing, the dynamics of the universe is very different and one would obtain instead a slowly rolling, inflating universe.  The exponential expansion is slowed down  to a power-law expansion. The density is no longer constant but keeps decreasing and can become arbitrarily small compared to its initial value. For an observer at a very late  time, the  effective vacuum energy density   is much smaller  than $\rho_{*}$.

This quantum dynamics of the Omega field can drive slow-roll inflation in the early universe.
It is convenient to define  slow-roll parameters as usual in terms of the fractional change in the Hubble parameter and its derivative:
\be
\ve_{H}:=-\frac{\dot{H}}{H^2}=-\frac{d\ln H}{Hdt} \quad \textrm{and} \quad \eta_{H}:=\frac{\dot \ve_{H}}{H\ve_{H}} =\frac{d\ln \ve_{H}}{Hdt} \, .
\ee
For our model we find
\be
\ve_{H} = \b^{2} \quad \textrm{and} \quad \eta_{H} =0 \, .
\ee
Slow-roll inflation that lasts long enough requires that $\ve_{H} \ll 1$ and $\h_{H} \ll 1$. 
Since $\b$ is small in the semiclassical approximation, these  conditions are satisfied. 
Note that the Omega field is not really a physical scalar but simply a mode of the metric in a particular gauge. Thus, the universe undergoes slow-roll inflation \textit{without } a fundamental scalar, driven entirely by vacuum energy through the nontrivial quantum dynamics of the Omega field. Our model thus  provides a two-dimensional realization of  Omega-driven inflation or `\textit{Omflation}'   in \cite{Dabholkar:2015qhk, Dabholkar:2015b}.  As it stands,  our model leads to an empty universe because it  simply keeps inflating. With matter fields, it would be possible to construct more realistic scenarios with graceful exit and primordial perturbations. 

In our two-dimensional model, dark energy in the present era is fundamentally no different than  the `dark energy' that drives omflation in the early universe. In particular, both decay slowly and the slow roll parameter would be given by $\ve_{H}$ as above. Of course, this is far from a realistic model for the history of the universe. We have not attempted to construct a graceful exit that can start a hot big bang. However, it raises the interesting possibility that dark energy today would also be slowly varying, perhaps with measurable variation. It would be interesting to construct a complete two-dimensional model of cosmology with these ingredients. We leave this problem for the future.

\subsection{Existence of the Quantum Theory \label{Existence}}
 
We have treated the timelike Liouville theory semiclassically. 
It would be very interesting if one can make sense of the quantum theory as a solvable model. It is well-known that the conformal factor of the metric  has a `wrong-sign' kinetic term \cite{Gibbons:1976ue}. For this reason, timelike Liouville theory  is a better model \cite{Polchinski:1989fn,Cooper:1991vg, DaCunha:2003fm, Martinec:2014uva} of  four-dimensional gravity rather than the much studied spacelike Liouville theory. 

Path integral quantization of timelike Liouville theory is complicated because the action is unbounded from below and the measure is not shift-invariant.  Fortunately, in  two dimensions,  the framework of conformal bootstrap \cite{Belavin:1984vu} offers another approach to defining the theory in terms of  the spectrum of operators together with  the two and the three point functions satisfying the bootstrap constraints.  All higher point functions  can in principle be constructed by gluing. There has been steady progress over the past decade in solving the timelike Liouville theory within this framework. In particular, the three point function which solves the Teschner relations \cite{Teschner:2001rv} was obtained in \cite{Zamolodchikov:2005fy, Kostov:2005kk}. This exact result can be reproduced from semiclassical computations  \cite{Harlow:2011ny} and using Coulomb gas methods \cite{Giribet:2011zx}. Moreover,  it  has recently been shown numerically that these three point functions solve  the full bootstrap constraints \cite{Ribault:2015sxa}.

Given this state of knowledge, one may hope  that all ingredients required for  defining  two-dimensional quantum gravity are in place. As a first step towards this goal,  a no-ghost theorem can be proven by studying the BRST cohomology of matter and ghosts coupled to timelike Liouville \cite{Bautista:2015c}. However, several subtleties remain as we summarize below. 
\begin{itemize}

\item The Weyl-invariant measure of the quantum gravity path integral is induced by the norm
\be
(\d \chi, \d \chi) := \int d^{2}x \sqrt{-h} \, e^{2\b\chi} \,  \d \chi(x) \, \d \chi(x) \, .
\ee
This measure on field space suppresses  quantum fluctuations from the regions  where $\chi$ is very negative. In  Liouville theory,  on the other hand, one uses a shift-invariant measure which has no such suppression. This raises the question whether Liouville theory  is the correct model for two-dimensional quantum gravity \cite{DHoker:1990ac}. In the case of spacelike Liouville,  the theory possesses a nontrivial duality symmetry under $b \rightarrow 1/b$. This suggests that the action contains the dual cosmological term in addition to the usual cosmological term
 \cite{Zamolodchikov:1995aa}.  This term  in the action  grows exponentially for very negative $\varphi$ and can suppress the unwanted quantum fluctuations. The situation is more complicated for timelike Liouville. Even though the theory possesses an analogous $\b \rightarrow 1/\b$  duality symmetry, the dual cosmological term is imaginary along the integration cycle that renders the path integral well-defined. Hence it is not clear if it suppresses the undesirable quantum fluctuations.
 
\item

Regions of very negative $\chi$ correspond to very short distances in the physical metric. If one wishes to suppress the quantum fluctuations by hand, then one would have to restrict the range of the field $\chi$ in the path integral.  This is a generic problem in defining a path integral over metrics.  A nonrenormalizable effective field theory such as gravity in four dimensions is well-defined  only up to some distance scale. When  the metric is dynamical, putting  a physical cutoff such as the Planck length  at short distances really requires putting a boundary in field space.  This is again a UV-problem.

\item
One can regard two-dimensional gravity as a model for super-critical bosonic string theory. 
From the target space perspective, the bosonic (super)-critical string theory contains a tachyon in the spectrum. The cosmological constant can be viewed as the expectation value of the tachyon field. In the target space, a field with a  tachyonic potential implies an instability of the vacuum, which is related to the $c=1$ barrier \cite{Seiberg:1990eb,Kutasov:1990sv}. Any nonzero expectation value away from the top of the potential will cause the system to roll away from the unstable extremum. The renormalization group beta function equations on the two-dimensional `worldsheet', which correspond to classical equations of motion in target space,  seem to support this conclusion \cite{Cooper:1991vg}. One may interpret the decaying value of the effective cosmological constant as rolling down the tachyon potential. It is possible that this renders even the semiclassical analysis unreliable as argued in \cite{Cooper:1991vg}. However, one must be cautious with such a naive interpretation. Timelike Liouville theory apparently makes sense  for all values of the cosmological constant. Moreover, the mass of the tachyon is comparable to the mass of other string modes. It is not meaningful to ignore these modes  in the beta function analysis. A satisfactory treatment  should take all modes into account, perhaps using closed string field theory. 

\end{itemize}

The presence of tachyons has  been identified with  a UV problem of the worldsheet theory corresponding to  too rapid asymptotic growth of  density of states  at high energy  \cite{Kutasov:1990sv}. One  possibility is that the tachyonic instability and the above mentioned problems are the analogs of the norenormalizability of higher dimensional quantum gravity. It is otherwise difficult to imagine an analog of the tachyonic instability in higher dimensions in `target  space' with mutliple universes joining and splitting. We hope to  return to these questions in \cite{Bautista:2015c}. 

In any case, the main lesson that we wish to abstract away  is the observation that when gravity is dynamical, various operators coupled to gravity such as the identity operator can have anomalous gravitational dressings, independent of the problem of nonrenormalizability. Even small values for these gravitational dressing can have observable effects in the cosmological setting when the universe undergoes exponential expansion with several $e$-foldings. 
%

\subsection{Local Form of the  On-Shell Quantum Momentum Tensor \label{Local1}}

If the fiducial metric is flat then it is possible to obtain  a local expression for the quantum  momentum tensor upon using the  equations of motion for the $\O$ field:
\be\label{eom-for-omega}
\nabla^2 \O-\half R_h +\frac{4\pi}{q} \m\, \b \,\cO_h^{\b}=0\, 
\ee
where we have used $q\b = 1-\b^{2}$. Using this equation with $R_{h}=0$, $S_{\m\n}$ in \eqref{Stensor} becomes
\bea
S_{\m\n}&=& -\frac{q}{4\pi \m\b}\left(\nabla_\m \nabla_\n- \half \h_{\m\n} \, \nabla \cdot \nabla \right)   \int dy \, 
G_{xy} \,  \nabla_{y}^2\O(y) \nonumber \\
&=& \frac{q}{4\pi \m\b}  \left(\nabla_\m \nabla_\n-\half \h_{\m\n}\nabla^2\right) \O(x)\, .
\eea\label{local-em}
The total quantum momentum tensor  can be obtained from the variation of the $\O$-dependent part of the quantum effective action \eqref{qaction1}:
\be
T^{\text{q}}_{\m\n}
= \frac{q^2}{2\pi}\left[ \frac{1}{\b q} \left( \nabla_\m\nabla_\n-\half \h_{\m\n} \nabla^2\right)\, \O(x)-\nabla_\m\O\nabla_\n\O +\half \h_{\m\n} (\nabla\O)^2\right] \, .
\ee
Interestingly, the original nonlocal expression has reduced to a local expression. 
It is instructive to compare this local expression  with the classical momentum tensor
\be
T^{\text{cl}}_{\m\n}=\frac{q^2}{2\pi}\left[ \left( \nabla_\m\nabla_\n- \half \h_{\m\n} \nabla^2\right)\, \O(x)-\nabla_\m\O\nabla_\n\O+\half \h_{\m\n} (\nabla\O)^2\right] \, .
\ee
Both tensors are properly traceless, and hence  $T_{+-} =0$. The $(++)$ components are
\bea
T^{\text{cl}}_{++} & = -\frac{q^{2}}{2\pi} \left[ (\partial_+\O)^2-\partial_+^2 \O \right] 
\, , \quad \b q = 1 \, ;\\
T^{\text{q}}_{++}&= -\frac{q^2}{2\pi} \left[ (\partial_+\O)^2-\frac{1}{\b q} \partial_+^2 \O\right] 
\, , \quad \b q = 1 - \b^{2}\, .
\eea
Imposing the Virasoro constraint corresponds to solving Einstein equations for spatially flat metrics near two dimensions. We see that the solution is given by 
\be
e^{\O(\t)} = e^{\O_{*}} (\frac{\t}{\t_{*}})^{\frac{-1}{\b q}} = e^{\O_{*}} (\frac{\t}{\t_{*}})^{\frac{2}{\g -2}} \, \,.
\ee
In the classical case we have $\b q =1$ and $\g =0$ whereas in the quantum case we have $\b q = 1-\b^{2}$ and $\g = 2\b^{2}$. 
With $a(\t) := e^{\O(\t)}$  and after writing the conformal time in terms of the comoving time the solution is in agreement with \eqref{density}.

 \subsection{Local Formulation with Auxiliary Fields \label{Local2}}

The nonlocal action \eqref{renormalized-integrated-cosmo-op} can be rewritten in a local form \cite{Nojiri:2007uq,Tsamis:2014hra} by introducing two auxiliary fields  $\Sigma(x)$ and $\Psi(x)$ with the  action 
\be
I_\L=
 -\m  \int d^2x \sqrt{-h} \,\left[ \, e^{2(1-\b^2) \Omega}\, e^{-2\b^2 \Sigma}\,+\, \Psi (2\,\nabla^2 \Sigma + R_{h})\,\right].
 \ee
 The equations of motion for the auxiliary fields are
\begin{eqnarray}
-\nabla^2 \Sigma&=& \half R_{h} \, , \label{eomauxiliary1} \\
-\nabla^2 \Psi&=&-\b^2 \, e^{2\Omega}\,e^{-2\b^2 (\Omega+\Sigma)}\, .\label{eomauxiliary2}
\end{eqnarray}
The first equation enforces  the field $\Sigma(x)$ to be the conformal factor of the fiducial metric $h_{\m\n}=e^{2\Sigma}\eta_{\m\n}$. After eliminating the auxiliary fields by using their equations of motion, we recover our  nonlocal action \eqref{renormalized-integrated-cosmo-op}.
The action is invariant under the Weyl transformation
\be
\Sigma\rightarrow \Sigma+\xi, \qquad \Omega\rightarrow\Omega-\xi, \qquad h_{\m\n}\rightarrow e^{2\xi}h_{\m\n}
\qquad \Psi\rightarrow \Psi\, .
\ee 
The local  momentum tensor resulting from this action is
\be\label{st with aux fields}
T^{\L}_{\m\n}= -\m \, \left[ h_{\m\n} ( e^{2\Omega}\, e^{-2\b^2 (\Omega+\Sigma)}-2\nabla \Psi \cdot \nabla \Sigma )
+4\nabla_{(\m}\Psi \nabla_{\n)}\Sigma + 2(\nabla_\m \nabla_\n -h_{\m\n} \nabla^2)\Psi \right]
\ee
which again reduces to \eqref{quantum-EMtensor} after using \eqref{eomauxiliary1} and \eqref{eomauxiliary2}.

\appendix 
\section{The Regularized Coincident Green Function}\label{appendixA}

We now compute the coincident Green function first using  short-time cutoff and then using dimensional regularization. Both methods are manifestly local and coordinate invariant\footnote{Other commonly used methods  use point-splitting \cite{Polchinski:1998rq} which is not manifestly covariant because of the choice of the direction used for point-splitting. One obtains the correct final answer by averaging over  directions.}. A common basic ingredient is  the $d$-dimensional heat kernel $K_{h}(x,y;s)$ satifying the heat equation 
\be
\left( \partial_s-\nabla_{h}^2\right)\, K_h(x,y;s)=\d(s)\,\d^{(d)}(x,y)\, 
\ee
with the initial condition
\be
K_h(x,y;0)= \d^{(d)}(x,y)\, .
\ee
In  flat space, the solution is given by
\be\label{flat-hk}
K_\d(x,y;s)=\frac{e^{-\frac{|y-x|^2}{4s}}}{(4\pi s)^{d/2}}\, .
\ee
Since the divergence of the coincident Green function comes from short distances,  it suffices to consider the adiabatic expansion of the heat kernel assuming small curvature
\be\label{d-leading-hk}
K_h(x,y;s)= \sqrt{\D_h(x,y)}\, \frac{e^{-\s(x,y)/2 s}}{(4\pi s)^{d/2}}\,\left[ 1+a_1(x,y) s+a_2(x,y) s^2+...\right] \, ,
\ee
 where the function $\s(x,y)$ is half the square of the geodesic distance between the two points and 
$\D(x,y)$ is the Van Vleck determinant
\be
\D_h(x,y) = \frac{\det \left[\partial_\m\partial_\n \,\s(x,y)\right]}{\sqrt{h(y)h(x)}}\, .
\ee
The adiabatic expansion parameter is effectively $s/L^2$, where $L^2$ is the typical radius of curvature.
In two dimensions, in  the conformal frame  one obtains in this approximation
\be\label{VanVleck-2d}
 \s(x,y)=\half e^{2\Sigma(x)}|x-y|^2 ,\qquad \D_h(x,y)= 1\, ,
\ee
where the exponential factor in \eqref{d-leading-hk} ensures that corrections are of order $\cO(\ve)$.
This reproduces the leading behavior of \eqref{2d-leading-hk}. 

For the diagonal heat kernel, the geodesic distance vanishes and the coefficients of the expansion $a_j(x)$ are the so-called Seeley-deWitt coefficients given in terms of  local curvature tensors. The Van Vleck determinant can be put to unity in Riemann normal coordinates. Therefore the short-time expansion of the diagonal heat kernel in $d$ dimensions reads
\be\label{diagonal-hk}
K_h(x,x;s)=\frac{1}{(4\pi s)^{d/2}}\,\left(1+a_1(x) s+a_2(x)s^2+...\right)\, .
\ee

\subsection{Short Proper Time Cutoff}

As discussed below \eqref{hk-reg-green} the coincident Green function can be regularized as \cite{DHoker:1988aa}
\be
G_h^\ve(x,x)=\int d^2y\sqrt{h}\,  \d_h^{(2)}(x, y) \, G_h(y,x)
= \int d^2y \,\sqrt{h}\, K_h(x, y;\ve) \, G_h(y, x)\, 
\ee
where the short-time $\ve$ effectively puts a cutoff on the distance between the two points. For small $\ve$  we need to keep only the leading term of the adiabatic expansion \eqref{d-leading-hk}. So in the conformal frame, using \eqref{VanVleck-2d} and
the explicit expression for the Green function \eqref{noncoincidentGreen}
\be
G_h^\ve(x,x)= -\frac{1}{4\pi}\,\int d^2y\sqrt{h}\, \frac{1}{4\pi \ve}\, {\exp{\left[-\frac{e^{2\Sigma(x)}\,|x-y|^2}{4 \ve}\right]}
}\, \ln(m^2 | y-x|^2)\,+\cO(\ve)\, .
\ee
The $\sqrt{h(y)}$ factor in the integrand is approximated by its value at point $x$ up to terms of higher order in $\ve$. Going to polar coordinates $r = |y-x|$
\be
G_h^\ve(x,x) = -\frac{1}{16\pi^2 \ve} \int \pi   d   r^{2} e^{2\Sigma(x)}\,\ln(m^2\, r^2) \,\exp{\left[ \frac{-e^{-2\Sigma(x)}\, r^2}{4\ve} \right]} \, .
\ee
A straightforward integration finally gives
\be\label{short-time-green}
G_h^\ve(x,x) = \frac{1}{2\pi}\Sigma(x) -\frac{1}{4\pi}\ln(4\, e^{-\g}\,m^2\,\ve)\, ,
\ee
where $\g$ is the Euler-Mascheroni constant.

\subsection{Dimensional Regularization}

The Green function is related to the heat kernel by
\be\label{green-hk}
G_h(x,y)=\int\limits_0^\infty ds\, K_h(x,y;s)\, .
\ee 
The coincident Green function is formally obtained by taking $x=y$ as the integral of the diagonal of the heat kernel \eqref{diagonal-hk}. Near two dimensions, only the first term of this integral has an ultraviolet logarithmic divergence  from the lower end of the integral. This can be regularized by continuing the integral to $d=2 +\e$ dimensions with $\e$ negative and small:
\be
G_h^d(x,x)\rightarrow G_h^\e(x,x)=\int\limits_0^\infty ds\, \frac{1}{(4\pi s)^{1+\frac{\e}{2}}}\, .
\ee
There is an infrared divergence from the upper end of the integral which can be regularized by introducing a mass term. Near two dimensions in the conformal frame the metric can be written as $h_{\m\n}=e^{2\Sigma_{h}}\d_{\m\n}$. Moreover, since only the first term in the adiabatic expansion \eqref{diagonal-hk} matters, we can take  $\Sigma_{h}$ to be a constant equal to its value at the point $x$. The infrared divergence can be regulated by considering the massive Green equation in a flat  metric $h_{\m\n}$ Weyl equivalent to $\d_{\m\n}$ by a \textit{constant} rescaling $e^{2\Sigma_{h}(x)}$:
\be\label{massive-green}
\left( -e^{-2\Sigma_{h}(x)}\, \d^{\m\n}\,\partial_{\m}\partial_{\n}  + m_{h}^{2}\right)G_{h}(x,y)  =e^{-2\Sigma_{h}(x)} \,{\d^{(2)}(x-y)}\, .
\ee 
Let  $m_{h}^{2} = m^{2} e^{-2\Sigma_{h}(x)}$. For fixed $m$ the massive Green equation is  Weyl invariant and the infrared regulator does no break Weyl invariance. The  regulated  Green function is  given by
\be
G_h^{\e}(x,x)=\frac{1}{(4\pi)^{1+\frac{\e}{2}}} \int\limits_0^\infty \frac{ds}{s^{1+\frac{\e}{2}}} \, {e^{-m^2 e^{-2\Sigma_{h}(x)}\,s}}\, \, 
\ee
for fixed $m$.  Following the discussion below \eqref{conformal-frame} this $m$ can be identified with the IR cutoff $1/R$ introduced in that section.The integral evaluates to
\bea
G_h^{\e}(x,x) &=&  \frac{1}{4\pi}\left[1 -\frac{\e}{2}\, \ln(\frac{4\pi e^{2\Sigma(x)}}{m^2})+ \cO(\e)\right]\,\G(-\frac{\e}{2}) \nn\\
&=&
-\frac{1}{2\pi \e}+\frac{\Sigma(x)}{2\pi}  - \frac{1}{4\pi}\ln( \frac{m^2 e^{\g}}{4\pi})+ \cO(\e)\, .
\eea

\section{Variation of the Scalar Green Function}\label{GreenVariation}

To compute the metric variation of the scalar Green function \eqref{general-variation-sigma} we vary the Green equation \eqref{green} to obtain
\be
\d\left( -\nabla^2\right)\, G_{yz}-\nabla^2\, \d G_{yz}= \half \, \d^{(2)}(y,z)\, h_{\m\n}\, \d h^{\m\n}
\ee
where the right-hand side follows from the variation of the $1/\sqrt{-h}$ factor of the delta function. Using the variation of the Laplacian
\be
\d \nabla^2=-\half \nabla_\a( h^{\m\n}\d h_{\m\n})\nabla^\a +(\nabla_\m \d h^{\m\n}) \nabla_\n + \d h^{\m\n} \nabla_\m\nabla_\n  \, 
\ee
we obtain a Poisson equation for $\d G_{yz}$ whose solution, after an integration by parts,  is given by
\begin{align}
\d G_{yz} 
=\int d w\, \d h^{\m\n}(w)\left(\half h_{\m\n} \nabla_\a\left( G_{yw}\nabla^\a G_{wz}\right)-
\nabla_\m G_{yw}\,\nabla_\n G_{wz}\right)
+\half h_{\m\n} \,\d h^{\m\n}(z)\, G_{yz}\,  
\end{align}
where all  derivatives are taken inside the integral are with respect to the variable $w$. 
The final expression for the functional  derivative is given by
\begin{align}
\frac{1}{\sqrt{-h}}\frac{\d G_{yz}}{\d h^{\m\n}(x)} = -
 \nabla^{x}_{(\m} G_{yx}\,\nabla^{x}_{\n)} G_{xz} + \half h_{\m\n}\, h^{\a\b}\,\nabla^{x}_\a G_{yx}\,\nabla^x_{\b} G_{xz} \, .
\end{align}
Note that this variation is traceless, as expected from the Weyl invariance of the Green equation.

\subsection*{Acknowledgments}

A major part of this work was conducted within the framework of  the ILP LABEX (ANR-10-LABX-63)  supported by French state funds managed by the Agence National de la Recherche within the Investissements d'Avenir programme under reference ANR-11-IDEX-0004-02,  and by the project QHNS in the program ANR Blanc SIMI5.  TB  thanks the HECAP group at the ICTP for hosting her visit.  AD  thanks the Department of Theoretical Physics at the Tata Institute of Fundamental Research where this work was initiated; and acknowledges the hospitality of the Aspen Center for Physics,  the Benasque Center  and the Theory Division at CERN  during part of this work. We thank Erik D'Hoker, Harold Erbin,  Diana L\'opez Nacir, Diego Mazzitelli, Raoul Santachiara, Leonardo Trombetta for useful discussions and  Paolo Creminelli, Jeff Harvey, and Ashoke Sen for comments on the draft.

We thank especially  La Grande Mosqu\'ee de Paris for the hookah and  th\'e à la menthe that stimulated this work. 

\bibliographystyle{JHEP}
\bibliography{weyl}
\end{document}